\documentstyle[11pt,aasms4]{article}
\begin{document}


\title{The variability analysis of PKS 2155-304}


\author{J.H. Fan}

\affil{
 Center for Astrophysics, Guangzhou Normal University, Guangzhou 510400,
 China, e-mail: jhfan@guangztc.edu.cn\\
Chinese Academy of Sciences-Peking
University Joint Beijing
Astrophysical Center (CAS-PKU.BAC), Beijing, China
}
\and
\author{R.G. Lin}
\affil{
 Center for Astrophysics, Guangzhou Normal University, Guangzhou 510400,
 China}

\date{Received <date>;accepted<date>}

\date{Received<data>;accepted<Oct. 1999>}

\begin{abstract}
 In this paper, the post-1977 photometric observations of PKS 2155-304 are
 compiled and used to discuss the variation periodicity. 
 Largest amplitude variations ( $\Delta U = 1^{m}.5$; 
 $\Delta B = 1^{m}.65$; $\Delta V = 1^{m}.85 $; $\Delta R = 1^{m}.25$; 
 $\Delta I = 1^{m}.14 $) and color indices ( $(B-V) = 0.30\pm 0.06$;
 $(U-B) = -0.72\pm 0.08$; $(B-R) = 0.62\pm 0.07$;
 $(V-R) = 0.32\pm 0.04$) are found.
 The Jurkevich's method and DCF (Discrete Correlation Function) 
 method indicate possible periods of 4.16-year and 7.0-year 
 in the V light curve.



\end{abstract}

\section{Introduction}
 BL Lac objects are a special subclass of active galactic nuclei (AGNs) 
 showing some extreme properties: rapid and large variability, high and 
 variable polarization, no or only weak emission lines in its classical 
 definition.

 BL Lac objects are variable not only in the optical band, but also in radio, 
 infrared,  X-ray,  and even $\gamma$-ray bands. Some BL Lac objects show that
 the spectral index changes with the brightness of the source (Bertaud et 
 al. 1973; Brown et al. 1989; Fan 1993), generally, the spectrum flattens when 
 the source brightens, but different phenomenon has also been found (Fan et
 al. 1999).

 The nature of AGNs is still an open problem; the study of AGNs variability 
 can  yield valuable information about their nature, and the implications for 
 quasars modeling are extremely important ( see Fan et al. 1998a).

 PKS 2155-304, the prototype of the X-ray selected BL Lac objects and
 TeV $\gamma$-ray emitter (Chadwick et al. 1999), is one 
 of the  brightest and the best studied objects. Its spectrum from 
 $\lambda 3600$ to $\lambda6800$ appears blue (B-V$<$0.1) and featureless 
 (Wade et al. 1979). A 0.17 redshift  was claimed from the probably detected
  weak [O III] emission  feature ( Charles et al.  1979), which was not 
 detected in  Miller \& McAlister (1983) observation.  Later, a redshift of 
 0.117 was  obtained from  several discrete absorption features (Bowyer 
 et al. 1984).   PKS 2155-304 varies at  all observation  frequencies and 
 is one of the most extensively studied  objects for both space-based  
 observations in UV and X-ray bands (Treves et al.  1989; Urry et al. 1993;  
 Pian et al. 1996; Giommi et al. 1998) and multiwavelength observations 
 (Pesce et al. 1997 ). Variation over a time scale of one day was observed
 (Miller \& Carini 1991) and that over a time scale of as short as 15 minutes
 is also reported by Paltani et al. (1997) in the optical band. 
  Differently  brightness-dependent spectrum properties are found ( see Miller
 \& McAlister 1983; Smith \& Sitko 1991; Urry et al. 1993; Courvoisier 
 et al. 1995; Xie et al. 1996; Paltani et al. 1997). 

 In this paper, we will investigate the periodicity in the light curve
 and discuss the variation as well. The paper has been arranged as
 follows: In section 2, the variations are presented and the periodicities 
 are searched, in section 3, some discussions and  a brief  conclusion are 
 given.

\section{Variation}
\subsection{Light curves}

 The optical data used here are from the literature: Brindle et al. (1986); 
 Carini \& Miller (1992); Courvoisier et al. (1995); Griffiths et al. (1979);
 Hamuy \& Maza (1987); Jannuzi et al. (1993);  Mead et al. (1990);  
 Miller \& McAlister (1983); Pesce et al. (1997); Smith \& Sitko (1991); 
 Treves et al. (1989); Urry et al. (1993);  Xie et al. (1996) and shown in 
 Fig. 1a-e. From the data,  the largest amplitude variabilities in UBVRI 
 bands are found:
 $\Delta U = 1^{m}.5 (11^{m}.87- 13^{m}.37)$; $\Delta B = 1^{m}.65 (12^{m}.55 - 14^{m}.20)$; 
 $\Delta V = 1^{m}.85 (12^{m}.27 - 14^{m}.13)$;
 $\Delta R = 1^{m}.25 (11^{m}.96 - 13^{m}.21)$; $\Delta I = 1^{m}.14 (11^{m}.55 - 12^{m}.69)$.
 and color indexes are found: 
 $(B-V) = 0.30\pm 0.06$ (N=140 pairs);
 $(U-B) = -0.72\pm 0.08$ (N=105 pairs); $(B-R) = 0.62\pm 0.07$ (N=90 pairs);
 $(V-R) = 0.32\pm 0.04$ (N=98 pairs), the uncertainty is 1$\sigma$  dispersion.

\subsection{Periodicity}

 The photometric observations of PKS 2155-304 indicate that it is  variable 
 on time scales ranging from days to years (Miller \& McAlister 1983).  
 Is there any periodicity in the  light curve?  To answer this question, 
 the Jurkevich (1971) method is used to  search for the periodicity in the 
 V light curve since there are more observations in this band.

 The Jurkevich method (Jurkevich 1971, also see Fan et al. 1998a) is based on 
 the expected mean square deviation and it is less inclined to generate 
 spurious periodicity than the Fourier analysis.  It tests a run of trial 
 periods around which the data are folded.  All data are assigned to $m$ 
 groups according to their phases around each trial period. The variance 
 $V_i^2$ for each group and the sum $V_m^2$ of all groups are computed.  
 If a trial period equals  the true one, then $V_m^2$ reaches its minimum. 
 So, a ``good'' period will give a much reduced variance relative to those 
 given by other false trial periods and with almost constant values.
 To show the significance of the trial periodicity, we adopted the $F$-test
 (see Press et al. 1992).

 When the Jerkevich method is used to V measurements,
 some results are obtained and shown in Fig. 2 ($m = 10$), which shows
 several minima corresponding to trial periods of less than 4.0-year and
 two broad minima corresponding to averaged periods of (4.16 $\pm$ 0.2)
 and (7.0 $\pm$ 0.16) years respectively.

 For the periods, which are smaller than 4.0-year, we found that the
 decrease of the $V_m^2$ is less than 3 times of the noise 
  suggesting that it is difficult for one to take them as real signatures of
  periods, i.e., those periods should be discussed with more observations.
  For the two broad minima, the $F$-test is used to check their reality.
  The significance level is 93.8$\%$ for the 4.16-year period and
  96.2$\%$ for the 7.0-year period.

\section{Discussion}
 
 PKS 2155-304 was observed more than 100 years ago. Griffiths et al. (1979)
 constructed the annually averaged B light curve up to the 1950's from Harvard
 photographic collection. But there are only a few  observations during the 
 period of 1950-1970.   The periodicity obtained here (see Fig. 2) are based 
 on  the post-1977 data.

  For comparison, we adopted the DCF (Discrete Correlation Function) method
  to the V measurements.  The DCF method,
 described in detail by
 Edelson \& Krolik (1988) (also see Fan et al. 1998b), 
  is intended for analyses
 of the correlation of two data set.
  This method can indicate the
 correlation of two variable temporal series with a time lag, and can
 be applied to the periodicity analysis of a unique temporal data set. If
 there is
 a period, $P$, in the lightcurve, then the DCF should show
 clearly whether the data set is correlated with itself with time lags of
 $\tau$ = 0 and $\tau$ = $P$. It can be done  as follows.

 Firstly, we have calculated the set of
 unbinned correlation (UDCF) between data points in the two data streams
 $a$ and $b$, i.e.

\begin{equation}
 {UDCF_{ij}}={\frac{ (a_{i}- \bar{a}) \times (b_{j}- \bar{b})}{\sqrt{\sigma_{a}^2 \times
 \sigma_{b}^2}}},
\end{equation}
 where $a_{i}$ and $ b_{j}$ are points in the data sets, $\bar{a}$ and $\bar{b}$
 are
 the average values of the data sets, and $\sigma_{a}$ and $\sigma_{b}$ are
 the corresponding standard deviations. Secondly, we have averaged the
 points sharing the same time lag by
 binning the $UDCF_{ij}$ in suitably sized time-bins in order to get the
 $DCF$ for each time lag $\tau$:

\begin{equation}
 {DCF(\tau)}=\frac{1}{M}\Sigma \;UDCF_{ij}(\tau),
\end{equation}
 where $M$ is the total number of pairs. The standard error for each bin is

\begin{equation}
\sigma (\tau) =\frac{1}{M-1} \{ \Sigma\; [ UDCF_{ij}-DCF(\tau)
]^{2} \}^{0.5}.
\end{equation}

The resulting DCF is shown in Fig. 3. Correlations are found with
time lags of (4.20 $\pm$ 0.2) and ( 7.31 $\pm$ 0.16) years. In addition,
there are signatures of correlation with time lags of less than 3.0 years.
If we consider the two minima in both the right and left
sides of the 7.0-year minimum, then we can say that the periods of 4.16 and
7.0-year found with Jerkevich method are consistent with the time lags
of 4.2-year and 7.3-year found with the DCF method.  
These two periods are used to simulate the light curve
(see the solid curve in Fig. 4).

It is clear that the solid curve does not fit the observations so well.
One of the reasons is that there are probable more than two periods
($\sim$ 4.2 and $\sim$ 7.0 years) in the light curve as the results
in Fig 2 and 3 indicate. Another reason is that the derived period is
not so significant as Press (1978) mentioned. Press argued that periods of
the order the third of the time span have a large probability to appear
if longer-term variations exist. 
The data used here have a time coverage of about 16.0 years, i.e., 
about 3 times the derived periods. Therefore, these are only tentative 
 and should be confirmed by independent work."

 From the data, the largest amplitude variations are found for UBVRI bands 
 with I and R  bands showing smaller amplitude variations. One of the 
 reasons is from  the fact  that there are fewer observations for those 
 two bands, another  reason is  perhaps from the effect of the host 
 galaxy, which affects the  two bands more seriously.
 
 In this paper, the post-1970 UBVRI data are compiled for 2155-304 to
 discuss the spectral index properties and to search for the periodicity.
 Possible periods of 4.16 and 7.0 years are found.

\acknowledgements{ We are grateful to the referee for his/her comments 
 and suggestions! This work is support by the National 
 Pan Deng Project of China and the National Natural Scientific Foundation 
 of China}

\newpage

Figure Captions\\

Fig. 1:  a: The long-term U light curve of PKS 2155-304;\\
b: The long-term B light curve of PKS 2155-304;\\
c: The long-term V light curve of PKS 2155-304;\\
d: The long-term R light curve of PKS 2155-304;\\
e: The long-term I light curve of PKS 2155-304.\\

Fig. 2:  Plot of $V_{m}^{2}$ vs. trial period, $P$, in years

Fig. 3:  DCF for the V band data. It shows that the V light curve
is self-correlated with time lags of 4.2 and 7.31 years. In addition,
there are also correlation with time lags of less than 4.0 years.

Fig. 4: The observed V light curve (filled points) and the simulated
V light curve (solid curve) with the periods of 4.16 and 7.0 considered.

\begin{figure}
\epsfxsize=18cm
$$
\epsfbox{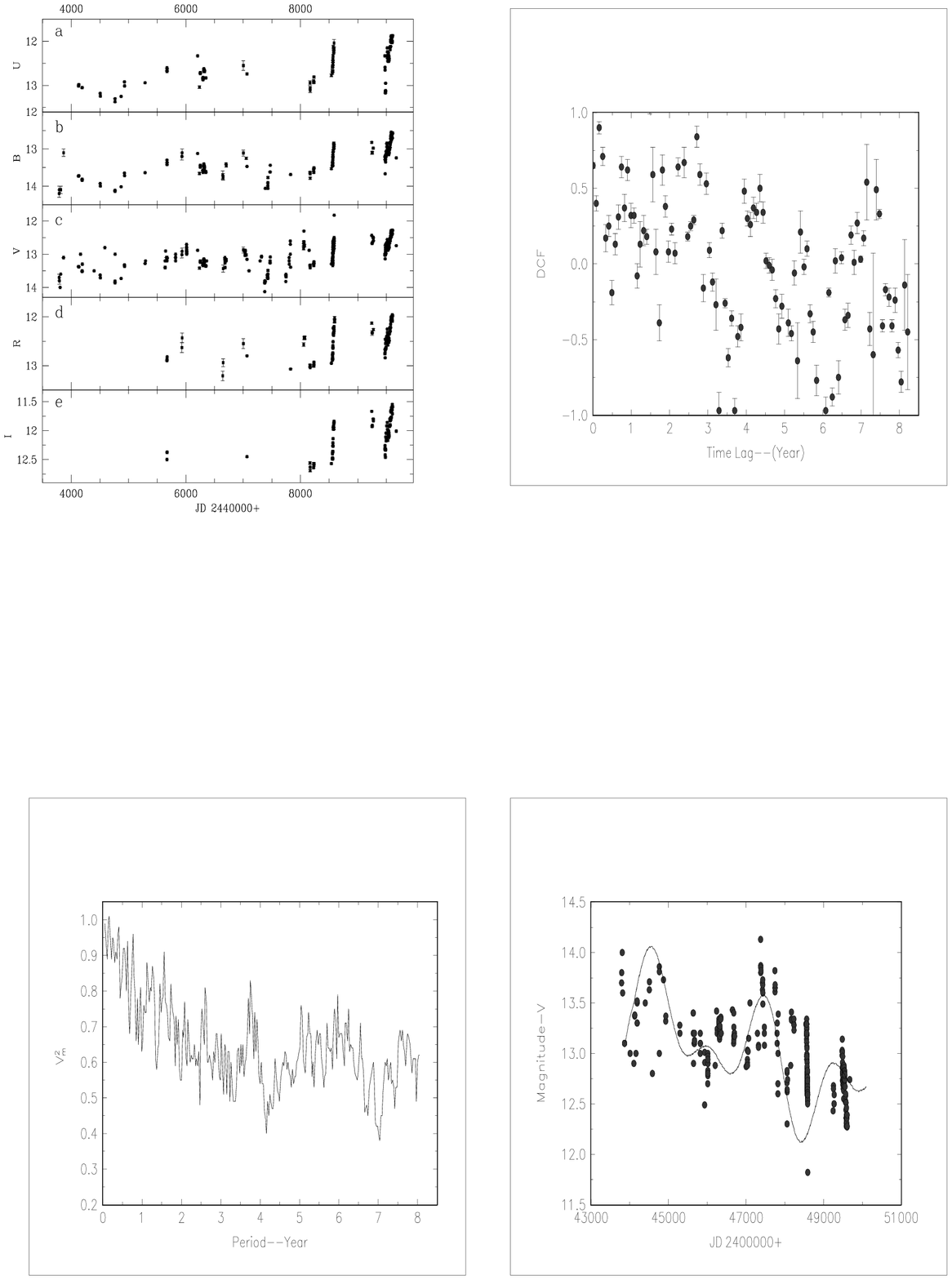}
$$
\caption {Fig1. Light curves of 2155-304; Fig 2. $V_{m}^{2}$ vs trial
period of 2155-304;
Fig 3. DCF analysis for 2155-304; Fig 4. The simulation light curve (solid
curve) and the 
 observation B light curve for 2155-304}
\label{fig:2}
\end{figure}

\end{document}